\RequirePackage{fix-cm}
\documentclass[smallcondensed]{svjour3}     
\smartqed  
\usepackage{amsmath}
\usepackage{graphicx}
\usepackage{lineno}
\usepackage{natbib}
\usepackage{upgreek}
\usepackage{psfrag}
\usepackage{subfig}
\usepackage{xcolor}
\usepackage[]{ar}
\definecolor{myblue}{RGB}{25, 96, 175}
\journalname{Boundary-Layer Meteorology}

\newcommand*\patchAmsMathEnvironmentForLineno[1]{%
\expandafter\let\csname old#1\expandafter\endcsname\csname #1\endcsname
\expandafter\let\csname oldend#1\expandafter\endcsname\csname end#1\endcsname
\renewenvironment{#1}%
{\linenomath\csname old#1\endcsname}%
{\csname oldend#1\endcsname\endlinenomath}}%
\newcommand*\patchBothAmsMathEnvironmentsForLineno[1]{%
\patchAmsMathEnvironmentForLineno{#1}%
\patchAmsMathEnvironmentForLineno{#1*}}%
\AtBeginDocument{%
\patchBothAmsMathEnvironmentsForLineno{equation}%
\patchBothAmsMathEnvironmentsForLineno{align}%
\patchBothAmsMathEnvironmentsForLineno{flalign}%
\patchBothAmsMathEnvironmentsForLineno{alignat}%
\patchBothAmsMathEnvironmentsForLineno{gather}%
\patchBothAmsMathEnvironmentsForLineno{multline}%
}
\newcommand{\rev}[1]{\textcolor{black}{#1}}

\begin{document}

\title{Vertical Coherence of Turbulence in the Atmospheric Surface Layer: Connecting the Hypotheses of Townsend and Davenport
}

\titlerunning{Vertical Coherence of Turbulence in the Atmospheric Surface Layer}        

\author{Dominik Krug \and
		Woutijn J. Baars \and
        Nicholas Hutchins \and
        Ivan Marusic}


\institute{D. Krug \at
             Physics of Fluids Group and Twente Max Planck Center, 
Department of Science and Technology, Mesa+ Institute,
 and J.M. Burgers Center for Fluid Dynamics, University of Twente, P.O Box 217, 7500 AE Enschede,  The Netherlands \\ \email{d.j.krug@utwente.nl} \and
             D. Krug \and W.J. Baars \and N. Hutchins \and I. Marusic \at
              Department of Mechanical Engineering, The University of Melbourne, Parkville, VIC 3010, Australia
}

\date{Received: DD Month YEAR / Accepted: DD Month YEAR}

\maketitle

\begin{abstract}

Statistical descriptions of coherent flow motions in the atmospheric boundary layer have many applications in the wind engineering community. For instance, the dynamical characteristics of large-scale motions in wall-turbulence play an important role in predicting the dynamical loads on buildings, or the fluctuations in the power distribution across wind farms. Davenport ({Quarterly Journal of the Royal Meteorological Society}, 1961, Vol. 372, 194-211) performed a seminal study on the subject and proposed a hypothesis that is still widely used to date. Here, we demonstrate how the empirical formulation of Davenport is consistent with the analysis of Baars {et al.} ({Journal of Fluid Mechanics}, 2017, Vol. 823, R2) in the spirit of Townsend's attached-eddy hypothesis in wall turbulence. We further study stratification effects based on two-point measurements of  atmospheric boundary-layer flow over the Utah salt flats. No self-similar scaling is observed in stable conditions, putting the application of  Davenport's framework, as well as the  attached eddy hypothesis, in question for this case. Data obtained under unstable conditions exhibit clear self-similar scaling and our analysis reveals a strong sensitivity of the statistical aspect ratio of coherent structures (defined as the ratio of streamwise and wall-normal extent) to the magnitude of the stability parameter.

\end{abstract}

\section{Introduction and Context}\label{sec:intro}

Coherence quantities of atmospheric surface-layer (ASL) turbulence are of great practical significance to the wind-engineering community as these are required for determining the dynamic action of  atmospheric turbulence on wind-sensitive structures, such as tall buildings, long span bridges \citep[][]{isyumov:2012a}, and wind turbines \citep{saranyasoontorn:2004a},  or in predicting peak-power distributions across wind farms \citep{sorensen:2007a,sorensen:2008a}.
Ground-breaking work on this subject is by Alan G. Davenport and can be found in \citet{Davenport1961}, \citet{davenport:1961phd}, \citet{simiu:1996a}, \citet{pasquill:1971a}, \citet{davenport:2002a} and \citet{baker:2007a}. The particular aspect we  focus on here is the degree and extent of the coherence of wind fluctuations in the vertical direction, which is  quantified via the linear coherence spectrum 
\begin{equation}\label{eq:LCS}
 \gamma^2\left(z,z_R;\lambda_x\right) \equiv \frac{\left\vert \left\langle X\left(z;\lambda_x\right) X^*\left(z_R;\lambda_x\right) \right\rangle \right\vert^2}{\left\langle \left\vert X\left(z;\lambda_x\right) \right\vert^2\right\rangle \left\langle \left\vert X\left(z_R;\lambda_x\right) \right\vert^2\right\rangle},
\end{equation}
where, $X\left(z;\lambda_x\right) = \mathcal{F}\left[\psi(z)\right]$ is the Fourier transform of some fluctuating quantity $\psi(z)$ and $\lambda_x$  is a streamwise wavelength. The vertical position is  $z$, and $z_R$  denotes the  reference position usually taken close to (or at) the surface. The asterisk * indicates the complex conjugate, $\langle \rangle$ denotes ensemble averaging and $\vert \vert$ is the modulus. 
 It is noted that the numerator equals the square of the  cross-spectrum magnitude, while the two energy spectra \rev{of $\psi(z_R)$ and $\psi(z)$} form the denominator.  Since $\gamma^2$ only incorporates the magnitude of the cross-spectrum, the value of $\gamma^2$ represents the maximum correlation for a specific scale $\lambda_x$.  Consequently, $\gamma^2$ equates to the fraction of common variance shared by $\chi(z_R)$ and $\chi(z)$ and we note that, by definition, $0 \leq \gamma^2 \leq 1$. As indicated in Eq. \ref{eq:LCS}, $\gamma^2$ generally is a function of the positions $z$ and $z_R$ and  the wavelength $\lambda_x$ (Note that we restrict the discussion to streamwise scales here as via Taylors hypothesis this direction is the most accessible experimentally).
 
 \citet{Davenport1961} hypothesised that the coherence is (i) a function of the ratio ${\Updelta} z/\lambda_x$ only, where $\Updelta z = z-z_R$,  and (ii) based on empirical observations proposed the functional form  
\begin{equation} \label{eq:DH}
\gamma^2_D= \textrm{exp}({-2k \Updelta z/\lambda_x}),
\end{equation}
where $k$ \rev{is a fit parameter to be determined from  experimental data.} Such a formulation is still widely used in the wind-engineering community to date \citep[e.g.][]{baker:2007a} and we will refer to it as Davenport's hypothesis.

During the same era as Davenport, A. A. Townsend made his impact in the field of turbulent shear flows \citep{marusic:2011a}, most notably with his attached-eddy hypothesis \citep{Townsend1976,Perry1982,Marusic2019}. A central tenet of the attached-eddy hypothesis states that  eddying motions in the logarithmic region of wall-bounded  flows are self-similar and that their size scales with their distance from the wall $z$. \rev{ In the context of the ASL, reference to the attached-eddy hypothesis has been made before, e.g. most recently by \citet{Li2018}}. 
Evidence in support of self-similarity and wall-scaling has been reported throughout the boundary layer community \citep[see for instance][]{jimenez:2012a,hwang:2015a,marusic:2017a} and most recently by \citet{Baars2017} who investigated the vertical coherence of the longitudinal velocity fluctuations relative to a reference very close to the wall.

Interestingly, it seems that regarding the coherence no cross-work exists between the two respective scientific communities to which Davenport (wind engineering) and Townsend (turbulent shear flow) belonged. To the authors' knowledge, only Davenport himself noted the early work of Townsend, as \citet[p. 209]{davenport:1961phd} states: \emph{``Some of the possible implications of this have been discussed by Townsend (1957)."} In this article we aim to connect the progress made in these communities regarding the understanding of the self-similar turbulent eddy structures in the ASL, as quantified by the coherence-diagnostic. In doing so, we will show that the geometrical self-similarity implied in Davenport's hypothesis is consistent with the attached-eddy hypothesis. Further, we will demonstrate that also the functional form given in Eq. \ref{eq:DH} agrees closely with a logarithmic dependence derived from the attached-eddy model \citep{Baars2017}.

We will start out by providing brief reviews of Townsend's and Davenport's hypotheses (Sect.\ref{sec:connect}) and demonstrate their conformity. Subsequently, we describe high-fidelity velocity and temperature data taken along the vertical direction in the atmospheric surface layer over smooth terrain (Sect.\ref{sec:data}). These data are used in Sect.\ref{sec:res} to infer the coherence statistics as a function of atmospheric stability. Throughout, we only employ the fluctuating components of the turbulence quantities; the streamwise (or longitudinal), spanwise and wall-normal velocity fluctuations are denoted by $u$, $v$ and $w$, respectively, with associated coordinates $x$, $y$ and $z$. Temperature fluctuations are denoted with $\theta$, its mean by $\Uptheta$.

\section{Connecting the Hypotheses of Townsend and Davenport}\label{sec:connect}
\subsection{Coherence Following Townsend's Attached-Eddy Hypothesis}\label{sec:townsend}
\citet{Townsend1976} envisioned that a wall-bounded shear flow encompasses a range of self-similar `attached eddies'. The terminology `attached' thereby implies that turbulence statistics scale with their distance from the wall, so-called $z$-scaling. The exact types of characteristic eddies and whether they are truly attached is of secondary importance. The AE description is applicable in the inertial region of the turbulent boundary layer (TBL), where the scales range from $\mathcal{O}(100)$ viscous units $\nu/U_\tau$ to the order of the boundary-layer thickness $\delta$. \rev{In practice, the inertial or `logarithmic' region of the ASL occupies the range from order of millimetres above the ground to $\mathcal{O}(100\mathrm{m})$ and therefore is highly relevant to all wind-engineering applications. } The ratio of the two length scales $\nu/U_\tau$ and $\delta$ forms the friction Reynolds number, $Re_\tau \equiv \delta U_\tau/\nu$, where $\nu$ is the kinematic viscosity and $U_\tau = \sqrt{\tau_0/\rho}$ is the friction velocity, with $\tau_0$ and $\rho$ being the wall-shear stress and fluid density, respectively.
Note that throughout the superscript `+' signifies normalization by `inner' scales $\nu/U_\tau$ and $U_\tau$. A  quantity analogous to the shear velocity, the wall conduction velocity, is given by $\Uptheta_\tau =  -\varepsilon \partial \Uptheta\vert_{z=0}/U_\tau$, $\varepsilon$ being the thermal diffusivity.

\begin{figure*}
\centering
\includegraphics[width = 0.999\textwidth]{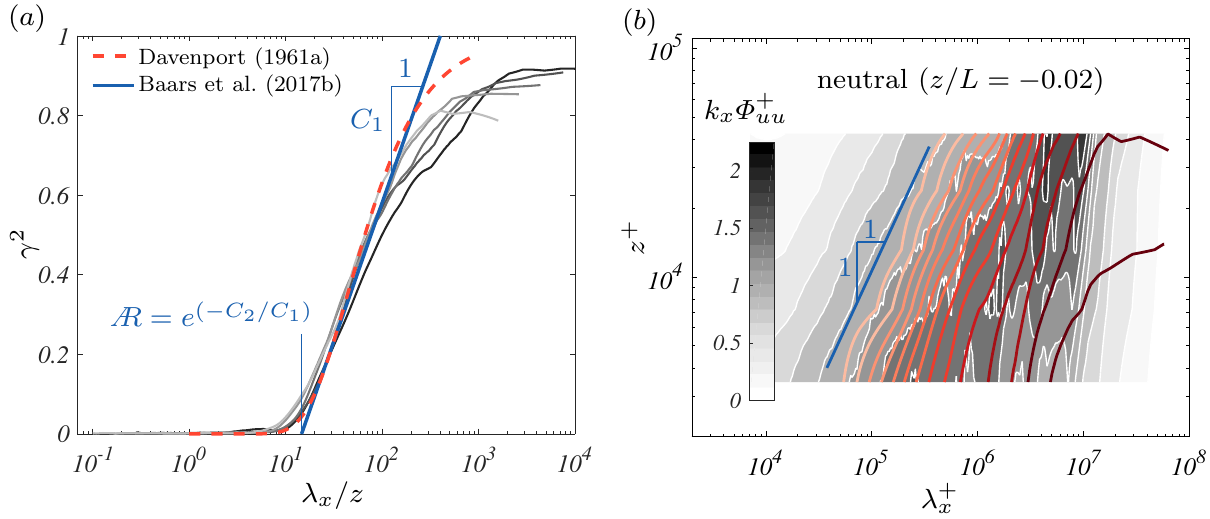}
\caption{(a) Linear coherence spectra for the neutral ASL \citep{Marusic2007}. Five spectra correspond to the five sonic-anemometers (darker shades of grey reflect increasing wall-normal distance). (b) Coherence spectra of sub-figure (a) presented as a spectrogram (red contours with levels 0.1:0.1:0.9 from left-to-right). The premultiplied streamwise energy spectrogram is shown with the underlying filled contours (grey contours with levels 0:2:0:2:1.8). The figure is adopted from \citet{Baars2017}.}
\label{fig:MH}
\end{figure*}

 Here, we review key results of  \citet{Baars2017}, who examined two-point measurements in the wall-normal direction for smooth-terrain and well-controlled flow conditions. \citet{Baars2017} considered measurements taken at the Utah SLTEST facility in ASL over salt flats at $Re_\tau \approx 1.4 \times 10^6$ \citep{Marusic2007}; further details of these types of experimental campaigns to study high-Reynolds-number wall-bounded turbulence can be found in the literature \citep{Metzger2007,Hutchins2012,wang:2016a,yang:2017a}. A wall-normal array of five sonic anemometers was employed, situated above a wall-shear-stress sensor. \rev{This unique set-up allowed them to investigate the  coupling between the outer-region turbulence and the near-wall footprint in the \emph{fluctuating friction velocity}. In most other cases only a near-wall velocity measurement is available, as will be considered later.
The coupling at different heights  was examined in spectral space using the linear coherence spectrum defined in Eq. \ref{eq:LCS}.}
 One coherence spectrum is obtained per velocity-pair $u(z_R)$--$u(z)$, where \rev{the height }$z$ ranges from $z^+ \approx 3\,500$ up to $z/\delta \approx 0.03Re_\tau$ \citep[corresponding to physical dimensions of $z = 2$ to 5\,m, ][]{Marusic2007}. Figure~\ref{fig:MH}a shows the five $\gamma^2$ spectra as a function of $\lambda_x/z$. Note that the streamwise wavelength $\lambda_x$ has been computed following $\lambda_x \equiv U(z)/f$, where $U(z)$ is the mean streamwise wind speed and $f$ is the temporal frequency.

A coherence spectrogram is formed by presenting the five individual coherence spectra as iso-contours of $\gamma^2$ in the $(\lambda_x,z)$ plane (Fig.\,\ref{fig:MH}b). The iso-contours increase in value with increasing wavelength, and the contours follow lines of constant $\lambda_x/z$ (slope of 1), reflecting the collapse of the individual spectra in Fig.~\ref{fig:MH}a. For reference, the energy spectrogram of the streamwise velocity fluctuations is shown with filled iso-contours underneath the $\gamma^2$ spectrogram. Energy is presented in premultiplied form $k_x\phi_{uu}/U^2_\tau$, where $\phi_{uu}$ is the one-sided power spectrum of $u$. Evidently, only a portion of the energy is statistically coherent with the near-wall measurement (the portion of the energy residing below non-zero contours of $\gamma^2$). 

\begin{figure*}
\centering
\includegraphics[width = 0.999\textwidth]{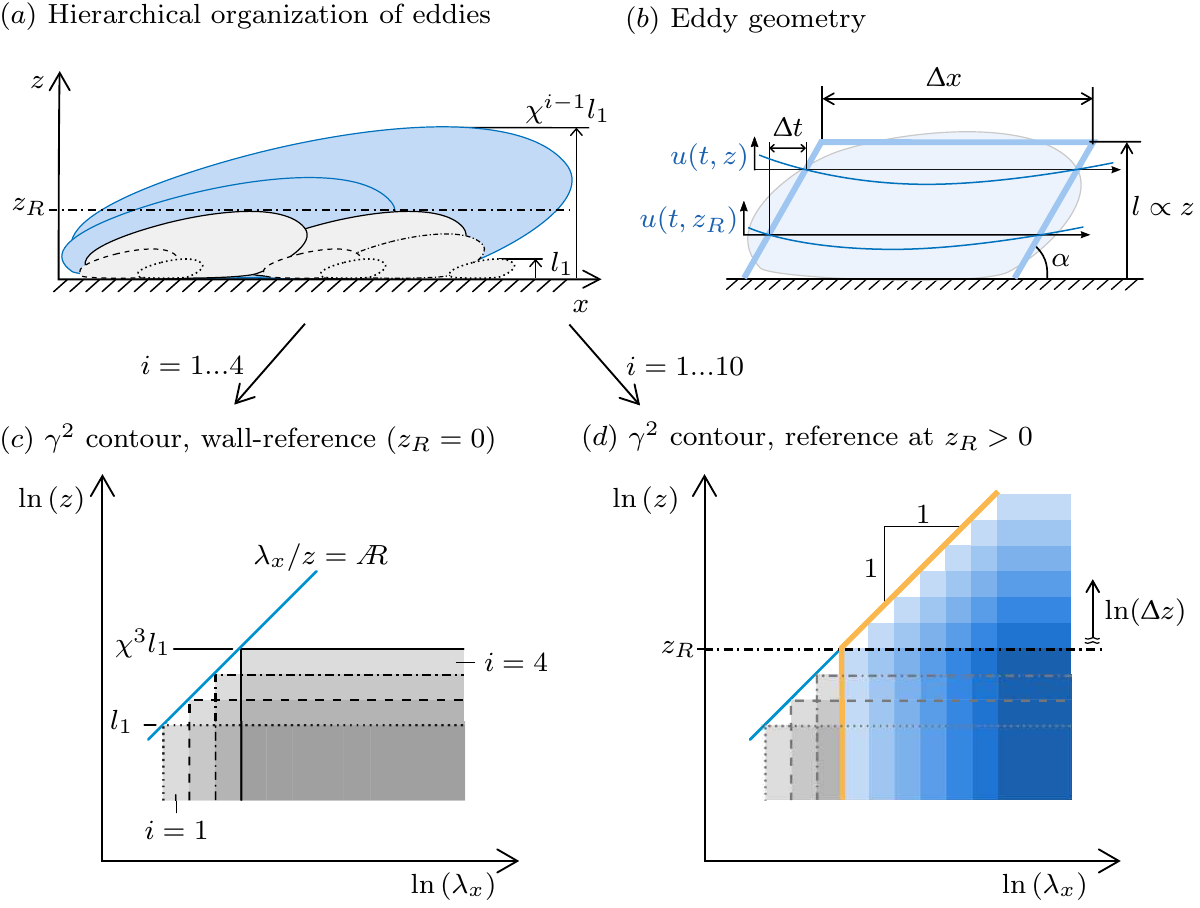}
\caption{(a) A hierarchical range of self-similar wall-attached turbulent eddy structures; for the purpose of this sketch, an arbitrary self-similar scaling factor of $\chi = 1.5$ is chosen. 
\rev{(b) Illustration of the eddy geometry. The thick blue line represents the parallelogram-like shape implied by the fact that the linear coherence considers the same wavelength at different $z$. The structure inclination angle $\alpha$ can equivalently be expressed as a phase shift $\Updelta t$ between signals at different wall-normal locations.}
(c) Filled iso-contours of $\gamma^2$ coherence of four discrete hierarchies (one grey shade per hierarchy $i = 1,...,4$), relative to a very near-wall reference position $z_R$. (d) Portion of the wall-reference-coherence spectrogram that survives when $z_R$ is placed at an arbitrary position in the logarithmic region (blue shaded contours).}
\label{fig:cohrecon}
\end{figure*}

Figure~\ref{fig:MH} demonstrates that only increasingly larger scales remain coherent with $z_R$ when $z$ moves upward. \rev{This trend is consistent with Townsend's attached-eddy hypothesis  in the formulation of \citet{Perry1982}, where a hierarchy of self-similar eddies is assumed as sketched in Fig.~\ref{fig:cohrecon}a. Each consecutive hierarchy is subject to an arbitrary scaling factor $\chi$. 
Figure~\ref{fig:cohrecon}b shows an idealized wall-attached turbulent structure with wall-normal extent $l$, streamwise scale $\Updelta x$ such that the aspect ratio $\AR \equiv \lambda_x/z$. 
When interpreting the coherence footprint of such a structure, it is important to recall that 
the coherence metric relates signal contributions at the same scale $\lambda_x \propto \Updelta x$. In the present application, this implies a parallelogram-like eddy structure as shown in Fig.~\ref{fig:cohrecon}b, for which $\Updelta x$ is the same at all $z$.  The widely observed phase difference in coherence metrics along the vertical direction \citep[e.g.][]{panofsky:1974a,Marusic2007,Chauhan2013,Liu2017,Salesky2018}, is equivalent to the inclination angle $\alpha$ of this structure as indicated in the figure. However, our definition of the aspect ratio only depends on $\Updelta x$ and $l$, i.e. the maximum wall-distance at which coherence at scale $\Updelta x$ is observed. The aspect ratio $\AR$ is hence solely determined by the magnitude of the coherence and insensitive to a phase shift (or inclination) between the turbulent fluctuations at $z$ and $z_R$. }


The implications of such a flow organization on the coherence spectrum with reference at the wall ($z_R=0$) is depicted in Fig.~\ref{fig:cohrecon}a. For each eddy hierarchy, there exists a minimum characteristic streamwise wavelength $\lambda_{x,i} \approx \Updelta x_i$ at which the structure appears coherent. 
Since eddies within the same hierarchy appear randomly in space (or time) \citep{Woodcock2015}, a non-zero coherence (and also turbulent energy) exists for scales larger than \rev{$\lambda_{x,i}$} within that hierarchy. Thus, for hierarchy $i$ with a wall-normal extent of $\chi^{(i-1)}l$, a non-zero contribution $\gamma^2_i$ to the coherence occurs in the region $z < \chi^{(i-1)}l$ and $\lambda_x > \AR \cdot \chi^{(i-1)}l$.
The magnitude of $\gamma^2_i$ may vary with $\lambda_x$ and $z$, but for simplicity we assume a uniform magnitude represented by a uniform grey-scale per $\gamma^2_i$ iso-contour in Fig.\,\ref{fig:cohrecon}b. Our conclusions, however, remain unaffected by eventual variations in  $\gamma^2_i$ as long as these remain self-similar across hierarchies as implied by the self-similarity of the underlying eddy field.
The full coherence spectrogram finally results from superposing the contributions of all hierarchies as reflected by increasing grey scales of the superposed transparent rectangles in Fig.~\ref{fig:cohrecon}b.
 Within a triangular region in $(\lambda_x,z)$ space, bounded by a minimum wall-normal height $z = l$, a constant $\lambda_x/z$ limit (at small wavelengths) and a constant $\lambda_x$ limit (large wavelengths), the $\gamma^2$ iso-contours align with lines of constant $\lambda_x/z$. Within this  region, the magnitude of $\gamma^2$ increases linearly with $\ln(\lambda_x)$ (for constant $z$) and decreases with $\ln(z)$ (for constant $\lambda_x$): a direct consequence of a geometrically self-similar structure. This implies that as a consequence of the attached-eddy hypothesis assumptions, the coherence magnitude within the self-similar region adheres to
\begin{eqnarray}
 \label{eq:baars}
 \gamma^2_{AE} = C_1\ln\left(\frac{\lambda_x}{z}\right) + C_2,
\end{eqnarray}
where $C_1$, $C_2$ are fit constants. The aspect ratio then follows from 
\begin{equation}
\AR= \frac{\lambda_x}{  z} \vert_{\gamma^2_{AE} =0} = \exp\left(\frac{-C_1}{C_2}\right).
\end{equation}
Based on laboratory data at $Re_\tau \approx 14\,000$, \citet{Baars2017} obtained  $C_1 \approx 0.302$ and $C_2 \approx -0.796$, which results in $\AR \approx 14$. These values were seen to be  consistent with numerical data at $Re_\tau \approx 2\,000$ \citep{Sillero2013} and the ASL data at $Re_\tau \approx 1.4 \times 10^6$ \citep{Marusic2007} (see the corresponding trend line in Fig.~\ref{fig:MH}).  All cases represent  the turbulent boundary layer under neutral stability conditions and were performed over smooth walls/terrain. 
Since (\ref{eq:baars}) holds for three orders of magnitude in $Re_\tau$, it can be concluded that a wall-attached, self-similar structure is indeed  ingrained in the $u$ velocity field. Similar conclusions for more isolated ranges in $Re_\tau$ were made by \citet{morrison:1969a} and \citet{bullock:1978a} for pipe flow and more recently by \citet{delalamo:2004a} for channel flow. Generally, all these results explicitly show the (wall-)coherent nature of the turbulence in the logarithmic region (at $\lambda_x > z\AR $). In particular, for all ASL type applications ($Re_\tau \sim 10^5$ to $10^6$), where the relative range of scales that are coherent will grow, these results reflect large-scale turbulent structures comprising significant lifetimes in the streamwise direction \citep{cantwell:1981a,robinson:1991a,hutchins:2007aa}.


Thus far, the coherence trend has been discussed only relative to the wall such that $z_R\to 0$. However, in typical tower micrometeorological studies, the reference measurement is taken at $z  \sim 1$\,m (which is well within the logarithmic region for typical atmospheric conditions). To illustrate how an off-wall position at height $z_R$ affects the idealized coherence trend in the attached eddy picture envisioned by Townsend, we increased the number of discrete hierarchies to 10 in Fig.~\ref{fig:cohrecon}c. For any given $z_R$, only the wall-attached turbulent structures that extend beyond $z_R$ are coherent with $z_R$ (their corresponding coherence contours are blue-shaded). The coherence trend above $z_R$ remains unaffected if only wall-attached structures are considered.

As a final remark, we point out that the above considerations made for the streamwise velocity component should also apply to the spanwise velocity field and temperature \citep{Perry1982,Krug2018}. 


\subsection{Comparing Davenport's Hypothesis to Townsend's}\label{sec:davenport}
\citet{Davenport1961} presented a trend in the wall-normal coherence of streamwise velocity fluctuations $u$ based on observations from typical tower micrometeorological data. It was evident from the data that both a decreasing wavelength $\lambda_x \equiv U/f$ and an increase in vertical separation $\Updelta z = z - z_R$ made the turbulent quantities less coherent. He hypothesized that for a given stability, the coherence should only depend on the ratio of $\Updelta z$ and $\lambda_x$. The implied geometrical self-similarity is obviously equivalent to the attached eddy framework discussed above if $\Updelta z \approx z $, which is approached either for $z_R \to 0$ or for $z \gg z_R$ in experiments.  The Davenport formulation, however, also entails self-similarity for any reference point, not just the wall, and therefore additionally encompasses also self-similarity of `detached' structures. 
Noting that the drop-off in coherence with increasing $\Updelta z/\lambda_x$ resembles an exponential decay, \citet{Davenport1961}  gave the following empirical expression
\begin{eqnarray}
 \label{eq:davenport}
 \gamma^2_D = \textrm{exp}\left({-2a\frac{\Updelta z}{\lambda_x}}\right),
\end{eqnarray}
where $a$ is a decay parameter. Here, a factor of two is added in the exponent compared to the original formulation, as Davenport proposed the relation for $\gamma$ (root-coherence) and we use $k \equiv 2a$ for brevity. Just as in other later works, we prefer  $\gamma^2$, since the squared coherence is proportional to the fraction of {energy} that is coherent over $\Updelta z$. 
With respect to the fitting constant in (\ref{eq:davenport}), Davenport initially quoted $a = 7.7$ for the `vertical coherence' ($\Updelta z$ separations) of the $u$ component in neutral conditions. Slightly updated values and extensions of the formulation to other velocity components and the temperature field have been reported in the ensuing literature \citep{pielke:1971a,davison:1976a,berman:1977a}. It is generally accepted that 
$a$ varies with surface-layer stability in the sense that it is small in strong convection and large in neutral or stable air, and we address this aspect in the following. A representation of (\ref{eq:davenport}) with $k = 23$ \citep{panofsky:1973c,naito:1974a} is included in Fig. \ref{fig:MH}a and is seen to match closely with (\ref{eq:baars}) and the data for neutral conditions.

\begin{figure*}
\centering
\includegraphics[width = 0.999\textwidth]{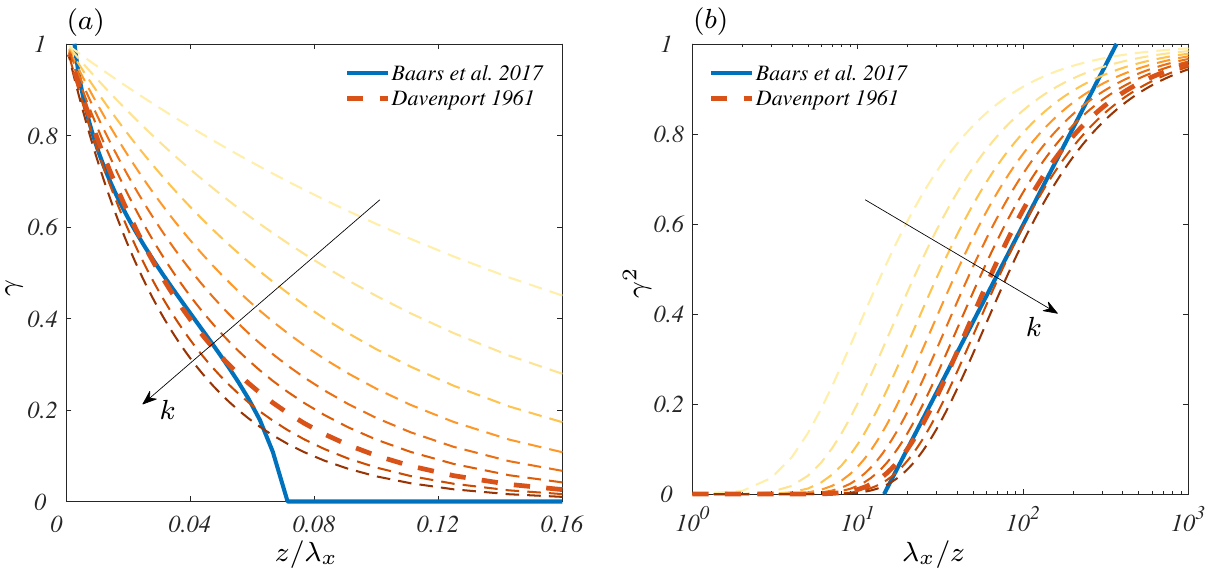}
\caption{ Parametrizations of the coherence spectrogram according to (\ref{eq:baars}) with $C_1 = 0.302$ and $C_2 = -0.796$ (solid blue line) and (\ref{eq:davenport}) with $k = 23$ (thick dashed red line). Panel (a) shows $\gamma$ vs. $z/\lambda_x$ on linear axes as commonly plotted in the wind-engineering community, in (b) $\gamma^2$ is plotted vs. $\lambda_x/z$ on semi-logarithmic axes. Additional thinner dashed lines in varying shades of red illustrate the effect of varying $k$ in (\ref{eq:davenport}) over a range $k=5$ to $k=39$ in increments of 3.}
\label{fig:param}
\end{figure*}

Since the introduction of (\ref{eq:davenport}) in 1961, many researchers have tested their data against Davenport's hypothesis. Studies range from research focusing on vertical coherence of various velocity components in tower micrometeorological data  \citep{Davenport1961,panofsky:1965a,pielke:1971a,naito:1974a,panofsky:1974a,brook:1975a,seginer:1978a,kanda:1978a,soucy:1982a,bowen:1983a,saranyasoontorn:2004a}, investigations including the lateral/spanwise coherence \citep{kristensen:1979a,ropelewski:1973a,panofsky:1975a,perry:1978a,kristensen:1979a2,kristensen:1981a,schlez:1998a}, the coherence of temperature fluctuations \citep{davison:1976a} and even meso-scale applications (typically in the horizontal directions) \citep{hanna:1992a,woods:2011a,vincent:2013a,larsen:2013a,mehrens:2016a}. 
Together, these measurements cover a great variety of terrain and topography. Here we wish to restrict the discussion to the effect of stability and limit the analysis  to the base case over smooth terrain, where high fidelity data are available from experiments at the Utah salt flats. 

Before we introduce this dataset (see Sect. \ref{sec:data}), we briefly consider stability effects in more detail. For this purpose, we have replotted  the parametrizations according to (\ref{eq:baars}) and (\ref{eq:davenport}) already included in Fig. \ref{fig:MH}a in Fig. \ref{fig:param} along with the Davenport parametrizations at varying $k$. As mentioned before, increasing stability corresponds to increasing $k$ and the linear plot of $\gamma$ vs. $z/\lambda_x$ in Fig. \ref{fig:param}a clearly illustratres how this leads to a faster decay of coherence. From the plot of $\gamma^2$ vs. $\lambda_x/z$ in Fig. \ref{fig:param}b it becomes apparent that a change in $k$ approximately leads to a horizontal shift of the curves. In the framework of \citet{Baars2017}, such a shift  implies a change in aspect ratio of the wall-attached structures  and we investigate this aspect in more detail below.

\section{Dataset} \label{sec:data}
The experimental dataset employed in this study has been recorded by \citet{Marusic2007},  \citet{Marusic2008} and was previously employed in a study of the neutral surface layer by \citet{Hutchins2012} and investigation of the stability dependence of the structure inclination angle by \citet{Chauhan2013}.  We refer to these papers for details beyond the short overview provided here.

The complete dataset consists of a continuous recording over several days  (26 May to 4 June 2005) at the Surface Layer Turbulence and Environmental Science Test (SLTEST) facility in the salt flats of western Utah. A measurement tower held nine logarithmically spaced sonic anemometers (Campbell Scientific CSAT3) at distances ranging from 1.42\,m to 25.69\,m above ground, which recorded all three components of velocity along with temperature. All measurements were synchronised and recorded at a sampling rate of 20Hz. In addition, there was a spanwise array at $z_s = 2.14$\,m above ground with 9 anemometers of the same type evenly spaced over 30\,m from the tower. Data from this array are only employed here to characterise the stability of the surface layer. 

Prior to further analysis, the data are corrected for wind direction and a de-trending procedure is applied \citep[see][for details]{Hutchins2012}. While the de-trending is necessary to remove slow temporal trends in the data, it inherently also compromises the coherence at very long scales, which needs to be kept in mind when interpreting the data. After data selection a total of 63 one hour long segments, corresponding to the dataset used in \citet{Chauhan2013}, remains. The stability of each segment is characterised using the Monin-Obukhov stability parameter $z_s/L$ with the Obukhov length scale
\begin{equation}
L =- \frac{\Uptheta U_\tau^3}{\kappa g \overline{w \theta}}.
\end{equation}
Here, the von K\'{a}rm\'{a}n constant $\kappa = 0.41$, $g$ is the gravity acceleration, and $U_\tau= (-\overline{uw})^{1/2}$ the friction velocity. All quantities are evaluated from an average of the total of 10 sonic anemometers at $z_s = 2.14$\,m. Most of our data lie in the unstable regime $z_s/L<0$ but a few data points also have $z_s/L>0$, which relates to stable conditions (downward heat flux).

Next, we will employ the SLTEST dataset to study how stability affects the self-similarity of coherent structures and how this changes the aspect ratio of the structures in the flow. We will be using the lowest measurement point as reference throughout, i.e. $z_R = 1.41$\,m from now on.

\section{Results}\label{sec:res}
\subsection{Stability Dependence of the Self-Similar Scaling}

\begin{figure*}
\centering
\includegraphics[width = 0.999\textwidth]{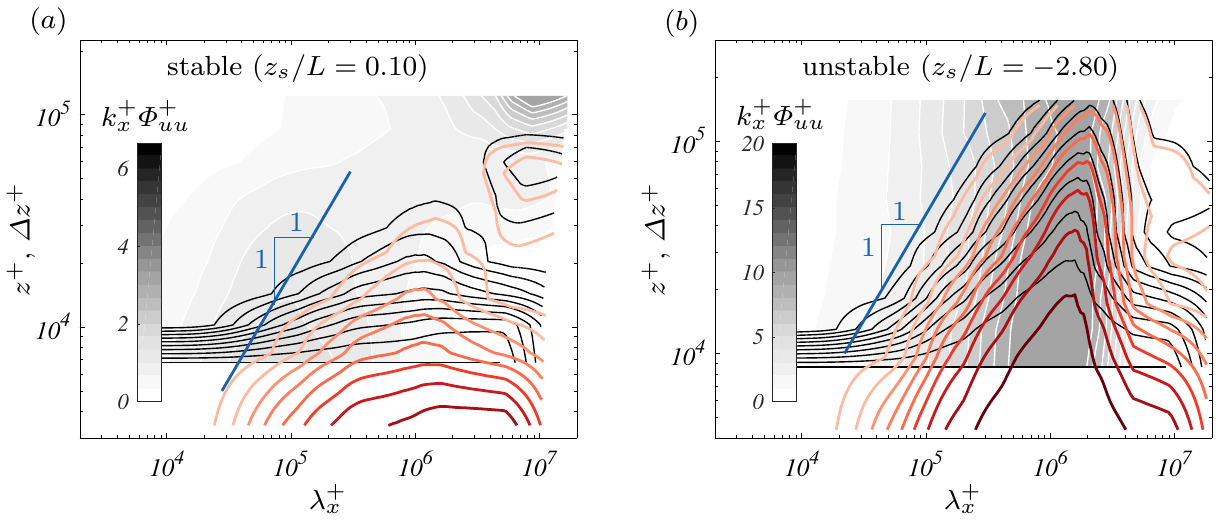}
\caption{Coherence spectrogram of the streamwise velocity component plotted vs. $z$ ( black lines) and $\Updelta z = z-z_R$ (shades of red) for a case with stable (a) and unstable (b) stratification.}
\label{fig:stabsample}
\end{figure*}

We start out by considering a representative case with stable stratification $z_s/L>0$ in Fig. \ref{fig:stabsample}a. Evidently,  the stable stratification has a significant effect on the spectral energy distribution of streamwise velocity fluctuations. Even more importantly, however, these changes are seen to also propagate to the coherence spectrogram. While the coherence levels are generally lower, the isolines are also seen to deviate notably from a slope of 1, which would be expected for self-similarity as discussed above (recall Fig. \ref{fig:MH}). This equally holds for plotting $\gamma^2$ vs. $z$ (as implied by the attached-eddy hypothesis), as well as vs. $\Updelta z$ following Davenport's hypothesis. The same observations could be made for the other stable data points (in total we have 11), other components of velocity and temperature (not shown here). Remarkably, this is already the case for relatively moderate values of $z_s/L$. In fact, the case at $z_s/L= 0.10$ shown in Fig. \ref{fig:stabsample} corresponds to the highest value in the dataset. Based on these findings, the application of both (\ref{eq:baars}) and (\ref{eq:davenport}) does not appear justified for stable stratifications and will not be pursued further here. 

The situation is different for unstable configurations where the heat flux is directed upward as can be seen from Fig. \ref{fig:stabsample}b. Also here, the turbulence field is significantly affected by buoyancy as evidenced by considerably higher fluctuation levels.  In contrast to the stable case however, the coherence isocontours in this case are observed to follow a slope of 1 for large enough $z^+$ consistent with the presence self-similar wall-attached structures. The fact that such a scaling is only obtained for sufficiently large $z^+$ is to be expected since, for small separations, also structures that are not attached to the wall will be coherent (trivially $\gamma^2 = 1$ at all scales $z = z_R$) and pure wall-scaling is only recovered once their influence has decayed. For large $\Updelta z$, there is no difference between plotting $\gamma^2$  as a function of $z$ and $\Updelta z$, consistent with the expectation at $\Updelta z\gg z_R$. A more interesting observation can be made at smaller $\Updelta z$, for which $\Updelta z\gg z_R$ does not hold.  Even for this range, the slope of the $\gamma$ isocontours is now approximately one, indicative of self-similar scaling,  all the way to the smallest vertical separation distance (noting that $\Updelta z = 0$ is not shown due to the logarithmic axis).  This means that self-similarity is now also observed where it was compromised by the contribution of non-attached structures when plotting with reference to the wall.
The extended scaling region is therefore, in fact, indicative of self-similarity of non-wall-attached structures  as suggested by the formulation of Davenport. Judging by the fact that the red lines for the ramp up in coherence are relatively straight, detached and wall-attached structures  have the same aspect ratios. As a side note to the discussion here, we would like to point out that the eventual decay of coherence for $\lambda_x>10^6$ is likely an artefact of the de-trending procedure since a similar effect was not observed in laboratory data \citep[see][]{Baars2017}.

\begin{figure*}
\centering
\includegraphics[width = 0.999\textwidth]{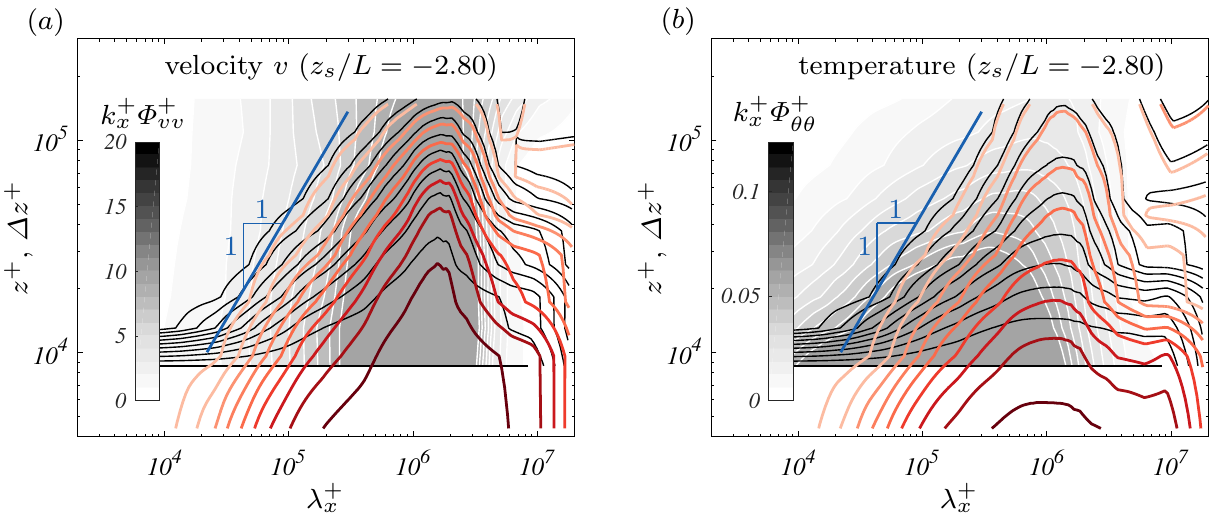}
\caption{Same as Fig. \ref{fig:stabsample} but for the spanwise velocity component $v$ (a) and temperature $\theta$ (b).}
\label{fig:compsample}
\end{figure*}

Our dataset allows us to investigate self-similar scaling also in the spanwise velocity component and the temperature fields. Such data is not commonly available in laboratory experiments and has therefore not been studied previously in this context. We plot the results corresponding to the unstable case in Fig. \ref{fig:stabsample}b but now  for $v$ and $\theta$ in Figs. \ref{fig:compsample}a and \ref{fig:compsample}b,  respectively. With maybe some minor limitations in the scaling vs. $z^+$ in the $v$-component, the situation is very similar to what was observed in the plot for the streamwise component in Fig. \ref{fig:stabsample}b. This provides evidence of self-similar scaling also in the spanwise velocity and the temperature field.  Especially the fact that the temperature field adheres to a similar geometrical organization appears remarkable in view of how different the energy spectrogram looks in this case. Compared to the velocity counterparts, the  relative contributions at high $\lambda_x$ as well as further away from the wall are considerably lower for the scalar spectrogram. 

\begin{figure*}
\centering
\includegraphics[width = 0.999\textwidth]{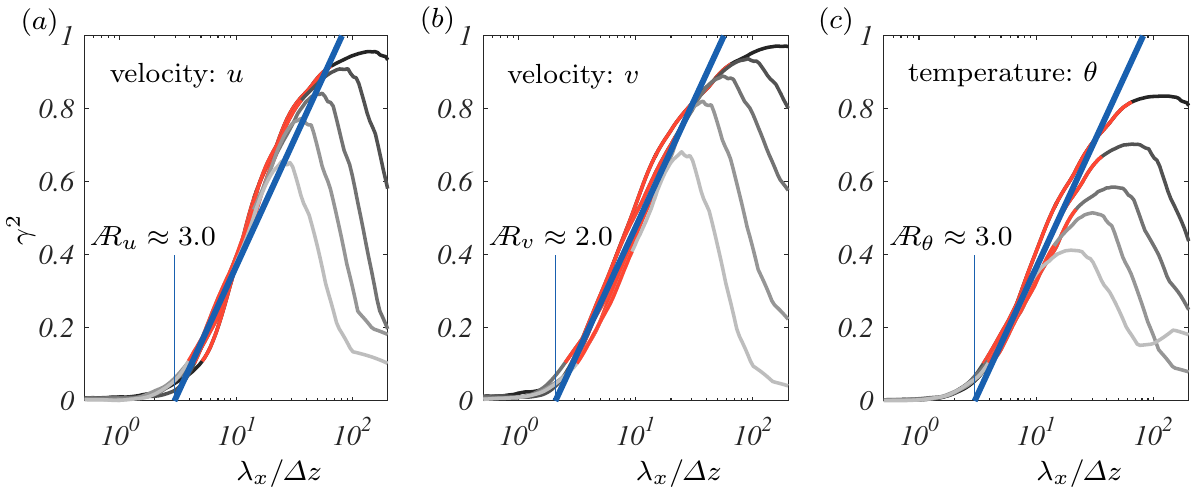}
\caption{\label{fig:selfsim} $\gamma^2$ in the range $1.58\,\textrm{m}\leq \Updelta z \leq 11.10\,\textrm{m} $  (with increasing $\Updelta z$ indicated by lighter shades of grey) for the streamwise (a) and spanwise (b) velocity components as well as temperature (c) for the case with $z_s/L = -2.80$. The blue line is a fit according to Eq. \ref{eq:baars} with $C_1=0.302$ fixed; the fitting region used here is bounded by $\gamma^2>0.1$ and $\lambda_x <100\,\textrm{m}$ and is indicated in red.  }
\end{figure*}

A more quantitative analysis is best achieved by replotting the data in the style of Fig. \ref{fig:MH}a. This is done in Fig. \ref{fig:selfsim} for the two velocity components and the temperature in the unstable case with $z_s/L = -2.80$. Note that here we use the normalization with $\Updelta z$ as this provides a  more extensive scaling region. Using $z$ instead provides similar results --- albeit with larger uncertainties. Plotting the results as presented in Fig. \ref{fig:selfsim} scrutinises the aspect of self-similarity, as for this case a collapse of curves at different $\Updelta z$ is expected. As can be seen from the figure, such a nominal collapse is indeed observed for all quantities plotted. Identifying the scaling region by the range of this collapse (marked in red in the figure) allows us to fit the expression (\ref{eq:baars}) to the data and to extract the statistical aspect ratio. These results are also included in Fig. \ref{fig:selfsim} and it is obvious that the resulting $\AR = 2.0$ to $\AR = 3.0$ is significantly lower than the value of $\AR \approx 14$ obtained under neutrally stable conditions. It therefore seems that unstable conditions drastically reduce the aspect ratio of coherent structures in the flow.

\begin{figure*}
\centering
\includegraphics[width = 0.999\textwidth]{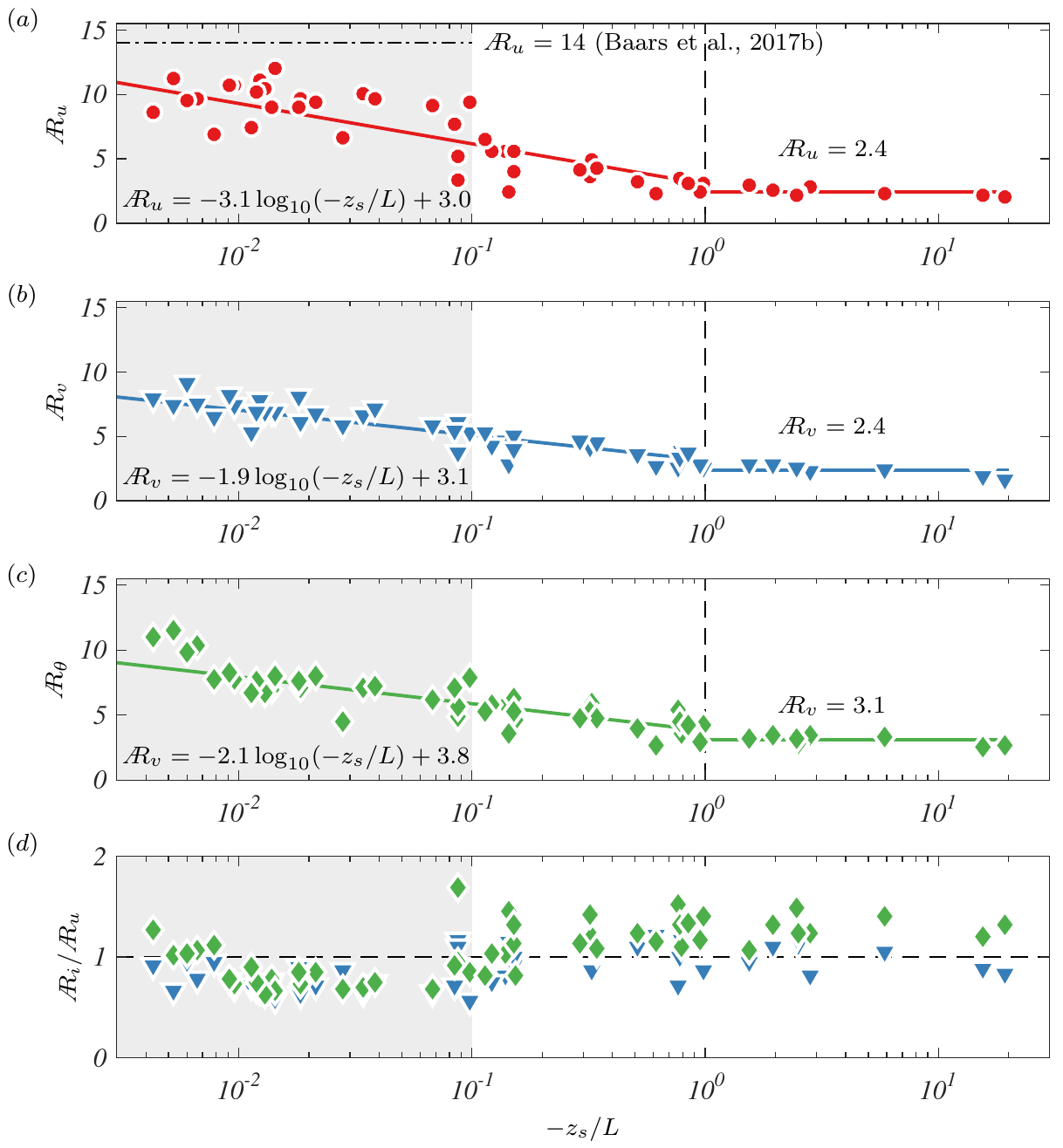}
\caption{\label{fig:AR} Results for the aspect ratio as a function of the stability parameter $z_s/L$ for velocity components $u$ (a) and $v$ (b) as well as temperature (c). Lines and the equations in (a-c) represent the results from fitting a linear relation for\rev{ $z_s/L<0.1$} and a constant for \rev{$z_s/L>0.1$}, respectively. Panel (d) compares the aspect ratios obtained from different quantities  by showing the ratios $\AR_v/\AR_u$ (blue triangles) and $\AR_\theta/\AR_u$ (green diamonds). The region shaded in grey denotes the   \rev{$-z_s/L <0.1$}, a range widely considered as near-neutral.   }
\end{figure*}

We systematically investigate this trend by applying the same fitting procedure to all unstable datasets available. The results obtained from doing so are presented for $u$, $v$, and $\theta$  in Fig. \ref{fig:AR}(a-c). In all three cases, there is a clear trend of decreasing $\AR$ with increasing magnitude of the stability parameter up to $-z_s/L \approx 1$. For even higher values of $-z_s/L$, the aspect ratio consistently attains approximately constant values in all three quantities in the range 2 to 3. It is worth highlighting how sensitive the aspect ratio is to even only weakly unstable conditions. Indeed, the data appears well represented by an --entirely empirical-- logarithmic fit for $-z_s/L <1$. The grey shaded region corresponds to $|z_s/L|<0.1$, which is a commonly used limit for approximately neutral conditions  \citep[e.g.][]{Hogstrom2002,Metzger2007,Hutchins2012}. $\AR$ varies by up to 50\% in this region, lending support to more stringent criteria such as $|z/L|<0.06$ applied in \citet{wang:2016a} at least for certain statistical quantities. We do note however, that the precise value of the stability parameter depends on the choice of the reference height $z$ (we use $z = 2.14$\,m) as $L$ is determined by wall quantities only. This choice is somewhat arbitrary for the multipoint statistics presented here and was motivated by data availability and consistency with previous studies in the present case. The dependence is linear however, and the values can therefore easily be transformed to other reference heights.

Finally, we briefly comment on the differences in the aspect ratios obtained from the three different quantities. We already established from Fig. \ref{fig:AR}(a-c) that the general trends are consistent across $u$, $v$ and $\theta$. In Fig. \ref{fig:AR}(d), we plot $\AR_v$  and $\AR_\theta $ with respect to $\AR_u$. In general, all ratios are close to 1. It will require further research to determine whether the slight deviations from 1, e.g. both $\AR_v/\AR_u$  and $\AR_\theta/ \AR_v$ tend to be slightly smaller 1 for $-z_s/L<1$, are statistically significant or simply owed to inaccuracies in determining $\AR$. Differences between the coherence decay in different velocity components have nevertheless also been reported in the wind engineering community where \citet{berman:1977a} and similarly \citet{pielke:1971a} report somewhat slower decay rates (lower $k$) for the $v$-component as compared to $u$.

\subsection{Relationship Between $\AR$ and $k$}

\begin{figure*}
\centering
\includegraphics[width = 0.999\textwidth]{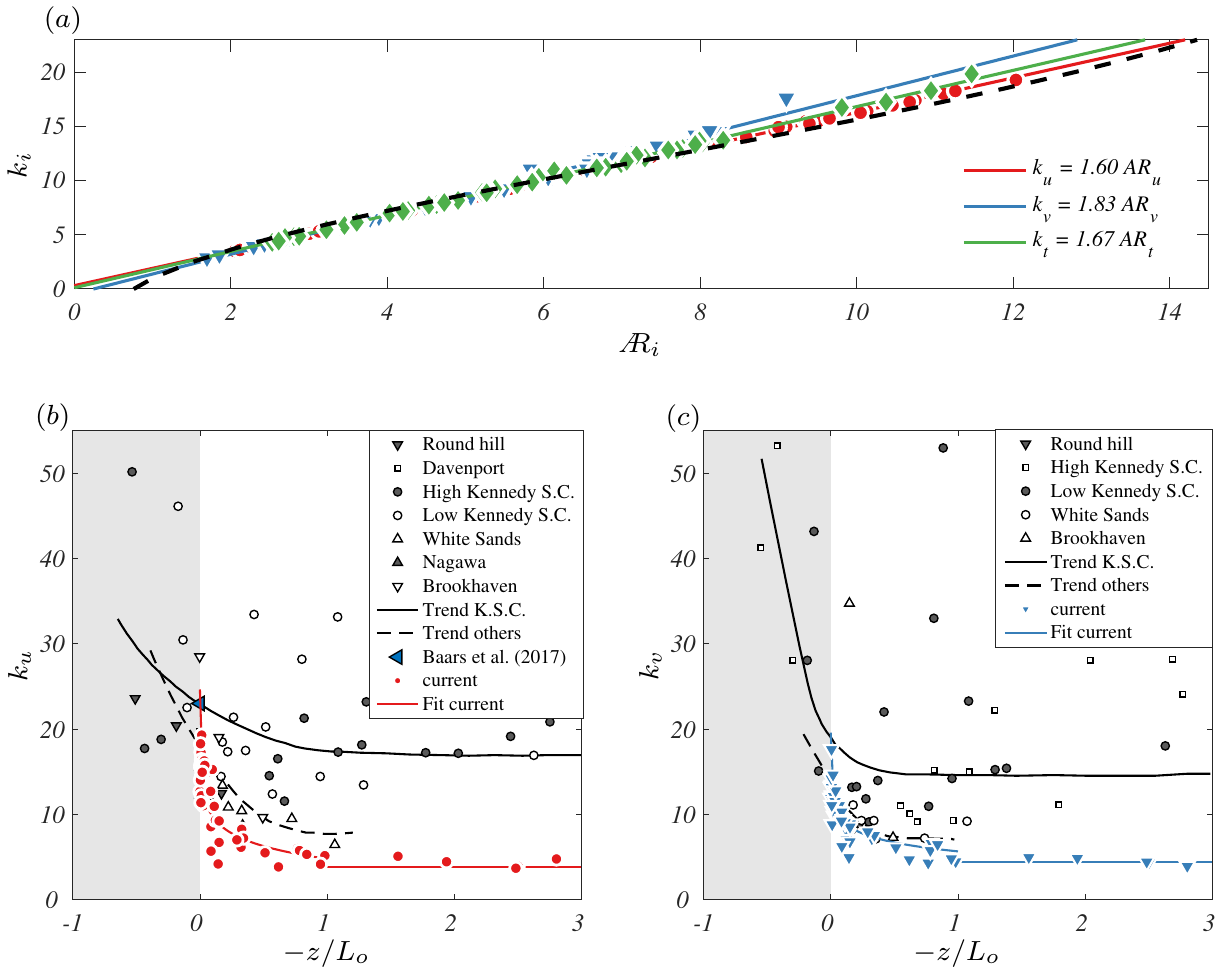}
\caption{(a) Fit results for $k_i$ vs. those for $\AR_i$ (same symbols as in Fig. \ref{fig:AR}), with $i = u,v,\theta$. Black dashed line represents the relation (\ref{eq:ana}) with $\xi = 20$, solid lines are linear fits to the data as labelled in the legend.
Panels (b) and \rev{(c)} respectively compare our results for $k_u$ and $k_v$ to literature values obtained from \citet{panofsky:1973c}. The fits to the current data shown here are obtained from the fits to $\AR$ in Fig. \ref{fig:AR}(a,b) along with the linear relations in panel (a) of this figure.
\label{fig:fit}}
\end{figure*}

The relationship between the statistical aspect ratio $\AR$ derived based on (\ref{eq:baars}) and the parameter $k$ in (\ref{eq:davenport}) is addressed in the following. An analytical relationship is readily derived by setting (\ref{eq:baars}) and (\ref{eq:davenport}) equal at a reference point $\xi \equiv \frac{\lambda_x}{z}\large|_{\textrm{ref}}$ and assuming $z = \Updelta z$. In this case, we obtain
\begin{equation}
k = -\xi/2\ln\left[C_1\ln(\xi)-\ln(\AR) \right],
\label{eq:ana}
\end{equation}
and from Fig. \ref{fig:selfsim}, $\xi = 20$ is a reasonable choice for the present data as it corresponds to $\gamma^2 \approx 0.5$. A representation of (\ref{eq:ana}) with this parameter is included in Fig. \ref{fig:fit}a along with results for $k_i$ obtained from fitting (\ref{eq:davenport}) to the same data points used to determine $\AR_i$ (here and in the following $i = u,v,\theta$). Overall, the agreement of (\ref{eq:ana}) with the data is very good and there is little scatter even between different components and the scalar. For the region of interest, the somewhat unwieldy relation  (\ref{eq:ana}) is very well approximated by a linear function and the corresponding linear dependence is included in Fig. \ref{fig:fit}a.
Figures \ref{fig:fit}b and \ref{fig:fit}c compare our results for $k_u$ and $k_v$ to values reported in the wind engineering literature and collected by \citet{panofsky:1973c}. In these figures the region corresponding to stable conditions is greyed out as our detailed analysis revealed a lack of self-similar scaling as implied by both (\ref{eq:baars}) and (\ref{eq:davenport}). 
The scatter in the literature data is considerable between different observation sites likely reflecting differences in topology. In this sense, our data from the SLTEST can be considered as a reference case with minimal roughness and topological effects. Consequently, the scatter in the present data is significantly lower. For both, $u$ and $v$, the new data points are largely consistent with the dashed lines labelled `Trend others', which is a trend line (by eye) for all measurements but those at the Kennedy Space Center (KSC) drawn by \citet{panofsky:1973c}. The data from the KSC (and the corresponding trend lines) lie at consistently and significantly higher values of $k_i$ than our results. Nonetheless, the fact that $k_i$ levels off at an approximately constant value beyond $-z_s/L=1$ is a feature that appears to be shared by all available datasets. Once again, we emphasise that the exact values of the stability parameter might change slightly depending on which reference height is chosen; unfortunately, information on the reference heights could not be retrieved for all literature datasets introducing an element of uncertainty.

\section{Concluding Remarks}\label{sec:concl}
The main findings of the present studies are summarised as follows:

\begin{itemize}
\item We have demonstrated that implications of Townsend's attached-eddy hypothesis for the coherence trend are consistent with Davenport's hypothesis. This applies to the geometrical self-similarity of wall-attached structures as well as to the fact that the empirically derived exponential decay in Davenport's formulation matches closely with a logarithmic expression that follows directly from the aspect of self-similarity.
\item The self-similarity implied by Davenport is even more comprehensive and also encompasses structures that are not attached to the wall. Evidence of such a self-similar behaviour could indeed be observed for the high $Re_\tau$ SLTEST data employed herein. 
\item The self-similarity assumptions/hypotheses do not seem to hold for stable data. Neither $z$- nor $\Updelta z$-scaling is observed in this case, which implies that there is no self-similarity for stable data. This is clearly observed in our data since we compute the coherence spectrum (a continuous function of scale with a finely frequency-discretized fast Fourier transform-approach). In the literature, the coherence is often computed at coarsely spaced frequency discretizations with fits based on a few data points only. Our results provide clear evidence that the stable ASL has no self-similar coherence and hypotheses of Townsend and Davenport should not be applied in this case.
\rev{We point out that it is predominantly the self-similarity aspect that fails in the stable regime while there is still non-zero coherence. The departure from self-similarity occurs far from extreme (`$z$-less') conditions at relatively weak stratification,  for which Monin--Obukhov similarity theory holds. 
 While we do not observe self-similarity for any of our stable data, it appears likely that for very weak stable stratification self-similarity may be recovered.  Unfortunately, we cannot determine such a threshold from the present dataset.} 
\item Consistent with expectations based on the attached-eddy hypothesis framework, self-similar scaling was not only observed for $u$, but also for spanwise velocity fluctuations $v$ and temperature fluctuations $\theta$.
\item Even relatively weak unstable stratifications drastically reduce the statistical aspect ratio $\AR$ for all quantities investigated here. Based on our results, we were able to parametrize this trend in terms of a logarithmic decay for $z_s/L<1$ and constant values for $z_s/L>1$. 
\item \rev{Generally, the trend of decreasing $\AR$ with decreasing stability is intuitively consistent with the fact that buoyancy supports the upward motion of structures from the wall. The question remains as to whether the nature of the structures themselves changes under very unstable conditions (e.g. towards convection cell-type motion) as may be suspected based on the change in trend for $\AR$ around $z_s/L =1$. A conclusive answer in this regard cannot be provided from the present analysis. It is, however, remarkable that the slope $C_1$ is largely insensitive to $z_s/L$ in our data. Physically, the parameter $C_1$ can be interpreted as a measure for the relative contribution of attached structures to the overall turbulence intensity. The fact that this quantity remains unaltered seems to indicate that, at least in the parameter range accessed here, the fundamental flow organization does not change significantly. This notion is substantiated by the observation that also self-similarity still holds in the unstable  regime.   }

\item We established a simple linear relation between \rev{the aspect ratio $\AR$ and  $k$ in the Davenport formulation, such that the fit parameter $k$ can be interpreted as an aspect ratio. }  Comparison of our results for the stability dependence of $k$ with the literature reveals significantly lower scatter for the high-fidelity ASL data over smooth-terrain presented here. As such, the present dataset serves as a base-case for the vertical coherence over any other type of terrain.
\end{itemize}

As a \rev{final} remark, we would like to point out that even though the present study is limited to vertical coherence, connecting Davenport's to Townsend's framework for this  case also places predictions and applications to horizontal coherence on a stronger footing.

\begin{acknowledgements}
The authors acknowledge financial support by the Australian Research Council  and by the University of Melbourne through the McKenzie fellowship program. We further thank Dr. Kapil Chauhan for making the de-trended data available to us.
\end{acknowledgements}

\bibliographystyle{spbasic}      
\bibliography{krug_blm}   

\end{document}